# THE N/O RATIO IN EARLY B-TYPE MAIN SEQUENCE STARS AS AN INDICATOR OF THEIR EVOLUTION

L. S. Lyubimkov

*It is shown that, in the case of early B-type main-sequence stars, of the three ratios N/C, C/O, and N/O which are regarded as indicators of stellar evolution, the ratio N/O is more reliable since it seems to be insensitive to overionization of the NII and OII ions. On the other hand, the N/C and C/O ratios, which include carbon, may contain systematic errors for stars with $T_{eff} > 18500$ K because of neglected overionization of CII ions. The ratio N/O is studied in the atmospheres of 46 early main-sequence B stars. These values of N/O are examined as functions of the effective temperature, age, rotation speed, and mass of the stars. Most early B-stars in the main sequence are found to have $[N/O] \approx 0$, which indicates that N/O varies little during the main sequence stage, and this result is independent of the basic parameters listed above. There are two explanations for the large number of stars with $[N/O] \approx 0$: it is predicted theoretically that for an initial rotation velocity $V_0 < 100$ km/s, N/O varies little toward the end of the main sequence stage ($[N/O] < 0.2$) and observations show that most early main-sequence B-stars do actually have low initial rotation velocities $V_0$. The few early main-sequence B-stars with higher $[N/O] = 0.4$-$0.8$ correspond to models with rotation velocities $V_0 = 200$-$300$ km/s. This conclusion is consistent with earlier data for stars with the same masses in a later stage of evolution: the AFG-supergiant and bright giant stage.*

Crimean Astrophysical Observatory, Russian Academy of Sciences, Russia; e-mail: lyub@craocrimea.ru

## 1. Introduction

During the main-sequence (MS) stage, the helium and nitrogen abundances in the atmospheres of early B-stars increase. This was first discovered by the author during 1970-1980, first for helium [1,2] and then for nitrogen [3]. It should be recalled that the MS stage is the first and longest lasting stage of stellar evolution, during which hydrogen burns in the core of a star. In early MS B-stars, hydrogen burns in the CNO cycle; at that time the abundances of He and N increase in the core of the star. The increase in the abundances of He and N discovered in the atmospheres of stars [1-3] seemed to indicate deep mixing, which should lead to transport of the products of the CNO cycle from the interior of the stars to their surfaces.

The observed increase in the abundances of He and N in B-stars during their evolution in the main sequence was unexpected for the theory of that time. An explanation for this phenomenon was found later, when theorists learned how to make computational models of rotating stars. It turned out that the initial rotation velocity $V_0$ is a parameter that is just as important for stellar evolution as the mass $M$ [4]. Models of stars with rotation showed that the changes in the atmospheric abundances of light elements owing to mixing induced by rotation are greater when the mass $M$ and rotation velocity $V_0$ are higher.

Mixing owing to rotation leads to higher abundances of helium and nitrogen and to a simultaneous reduction in the carbon content. (Here and in the following we consider the most widespread isotopes of these elements, $^4$He, $^{14}$N, and $^{12}$C). An anticorrelation between the abundances of N and C in A-, F-, and G-supergiants, which are descendants of early main sequence B-stars, has long been known. It has recently been shown [5] that the observed anticorrelation between N and C in these stars primarily reflects the dependence of the anomalies in N and C on the initial rotation velocity $V_0$: when $V_0$ is higher, the excess of N and deficit of C are larger. This confirms the importance of the parameter $V_0$ for a correct understanding of the evolutionary changes in the observed chemical composition of stars.

It is known [6] that the ratios of the abundances of three elements that participate in the CNO cycle, specifically N/C, C/O, and N/O, are among the most sensitive indicators of the evolution of stars with medium and high masses. The ratios N/C and N/O for models with masses M ranging from $15\,M_\odot$ to $1.7\,M_\odot$ (where $M_\odot$ is the sun's mass) and rotation velocities $V_0$ ranging from 0 to 0.95 times the critical velocity have been calculated in detail [7].

This paper is an analysis of the ratio N/O for groups of early main sequence B-stars. The ratios N/C and C/O, which involve carbon, are not considered, since their values for these stars may be systematically distorted because of neglected overionization (superionization) of CII ions (see below). The dependence of N/O on the age, rotation velocity, and mass of the stars is examined. The results are compared with the earlier calculations [7] for rotating star models.

## 2. How real are the determinations of N/C, C/O, and N/O for early B-stars?

This paper is based on previous studies of MS B-stars which were published between 2000 and 2008 by Lyubimkov, et al. [8-12]. High resolution spectra for more than 100 MS B-stars were obtained [8] at two observatories, the Crimean Astrophysical Observatory and the MacDonald Observatory of the University of Texas. The fundamental parameters of 107 of the stars in this list were subsequently determined [9], including the effective temperature $T_{eff}$, acceleration log$g$ of gravity in the atmosphere, radius $R$, luminosity $L$, and age $t$, as well as the relative age **t/t$_{MS}$**, where $t_{MS}$ is the lifetime of a star with this mass $M$ in the main sequence stage. The helium abundance for 102 of the stars was determined [10] using a non-LTE (local thermodynamic equilibrium) analysis of the HeI lines. A non-LTE analysis of the magnesium content has been carried out for 52 of the stars [11] and a non-LTE analysis of the abundances of C, N, and O for 57 of the stars [12].

Later, the same authors chose [13] 22 B-stars with masses $M = 5-11 M_\odot$ that had not finished evolving and, after re-evaluating $T_{eff}$ and log$g$ for these stars using an improved method, found their abundances of C, N, and O. These abundances were close to the solar values, which confirmed that the initial chemical composition of the atmospheres of young MS B-stars are the same as those of the sun on the average. Only carbon manifested a small deficit, as others have found.

In the context of this study, Ref. 14 is of particular interest: it deals with possible overionization of CII, NII, and OII ions in the atmospheres of early B- and late O-stars. Recall that the abundances of C, N, and O for early B-stars are determined from CII, NII, and OII lines. Here the term "overionization" means that photoionization of CII, NII, and OII ions by ultraviolet radiation can be substantially higher in real atmospheres than the calculated level based on standard atmospheric models. These models have been used up to now for analyzing the abundances of C, N, and O in early B-stars.

It has been shown [14] that overionization becomes noticeable at temperatures $T_{eff}>18500$ K for CII lines and $T_{eff}>26000$ K for NII and OII lines. By choosing (from Ref. 13) stars with $T_{eff}<18500$ K for estimating the abundance of C and with $T_{eff}<25000$ K for estimating the abundances of N and O, it was possible to obtain initial abundances of C, N, and O in MS B-stars that were undistorted by overionization. These values are the same as the solar abundances to within the limits of error. This can be seen in Table 1, which shows the average initial abundances $\log\varepsilon(C)$, $\log\varepsilon(N)$, and $\log\varepsilon(O)$ for MS B-stars found in Ref. 14, together with recent estimates of the abundances of C, N, and O for the sun [15,16]. The latter were obtained using nonstationary hydrodynamic 3D models of the solar atmosphere. We note that the elemental abundances here and in the following are given in the standard logarithmic scale, where for hydrogen it is assumed that $\log\varepsilon(H)=12.00$.

TABLE 1. Average Initial Abundances of C, N, and O in the Atmospheres of MS B-Stars in the Vicinity of the Sun [14]. Data on the Abundances of C, N, and O in the Sun's Atmosphere [15,16] are Shown for Comparison.

| Element | log$\varepsilon$ MS B-stars | log$\varepsilon$ Sun |
|---|---|---|
| C | 8.46±0.09 | 8.43±0.05 [15] <br> 8.50±0.06 [16] |
| N | 7.78±0.09 | 7.83±0.05 [15] <br> 7.86±0.12 [16] |
| O | 8.72±0.12 | 8.69±0.05 [15] <br> 8.76±0.07 [16] |

Since, according to Ref. 14, for temperatures $T_{eff}>18500$ K the derived abundance of C is low (by up to 0.2 dex) because overionization is neglected, while the abundances of N and O up to temperatures $T_{eff}\approx 26000$ K are undistorted, in the range $T_{eff}\approx 18500-26000$ K the N/C ratio should be higher and the C/O ratio should be lower than the actual value. This implies that for early MS B-stars (which correspond to temperatures $T_{eff}>18000$ K), the published values of N/C and C/O may be correspondingly elevated or reduced (by up to 0.2 dex).

The situation is different for the ratio N/O. In the case of B-stars with temperatures $T_{eff}<26000$ K, nitrogen and oxygen are both insensitive [14] to overionization, so for these elements a determination of their ratio N/O can be regarded as fully reliable. In addition, for stars with $T_{eff}>26000$ K, superionization has roughly the same effect on NII and OII, i.e., the abundances of N and O are simultaneously lowered and this should have little effect on the ratio N/O.

It is interesting to note that near $T_{eff}\approx 26000$ K a maximum is observed in the distribution of the equivalent widths W of the NII and OII lines with respect to $T_{eff}$ (see, for example, Figs. 2 and 3 of Ref. 13). However, the ratio of the equivalent widths, W(NII line)/W(OII line) manifests no maximum or minimum in near $T_{eff}\approx 26000$ K; on the other hand, this ratio varies monotonically over a wide range from $T_{eff}=14000-32000$ K.

The above discussion shows that the ratio N/O is to be preferred in studies of the ratios N/C, C/O, and N/O for early B-stars since N/C and C/O, which involve carbon, may be systematically distorted by overionization of CII ions. Based on this conclusion, the author avoids using N/C and C/O. This paper is a study of the ratio N/O for a rather large group of early MS B-stars.

**3. The list of stars to be studied**

The B-stars selected for this analysis from Ref. 12 are listed in Tables 2 and 3. Since our problem was to determine the ratio N/O, the stars for which the oxygen abundance was not determined [12] have been eliminated from this list. In addition, several stars that were spectrally binary or variable according to the SIMBAD data base (http://simbad.u-strasbg.fr/simbad/sim-fid) have also been eliminated.

Table 2 lists the parameters of 40 early B-stars with effective temperatures $T_{eff}<26000$ Ê. As noted above, for stars with these temperatures $T_{eff}$ the N and O abundances (and, thereby, the ratios N/O) are not distorted by overionization. Besides the HR number, in Table 2 the following parameters are listed for each star: effective temperature $T_{eff}$, logarithm of the acceleration of gravity $\log g$, mass in terms of the sun's mass $M/M_\odot$, relative age $t/t_{MS}$, and projection of the rotation velocity on the line of sight $V\sin i$. Note that the values of $V\sin i$ have been found [10] using a non-LTE analysis of the profiles of six HeI lines.

TABLE 2. Parameters of 40 B-Stars with Temperatures $T_{eff}$ < 26000 K

| HR | $T_{eff}$ K | log$g$ | $M/M_\odot$ | $t/t_{MS}$ | $V$sin$i$ km/s | log(N/O) | [N/O] |
|---|---|---|---|---|---|---|---|
| 1 | 2 | 3 | 4 | 5 | 6 | 7 | 8 |
| 38 | 19220 | 3.83 | 7.6 | 0.76 | 11 | -0.86 | +0.08 |
| 1072 | 22300 | 3.81 | 9.8 | 0.76 | 39 | -0.22 | +0.72 |
| 1074 | 25700 | 4.22 | 10.5 | 0.06 | 97 | -1.09 | -0.15 |
| 1595 | 22500 | 4.17 | 8.4 | 0.21 | 6 | -0.83 | +0.11 |
| 1617 | 18570 | 3.87 | 7.1 | 0.70 | 46 | -1.00 | -0.06 |
| 1640 | 19750 | 3.81 | 8.0 | 0.76 | 54 | -1.05 | -0.11 |
| 1731 | 17900 | 3.98 | 6.2 | 0.60 | 11 | -0.72 | +0.22 |
| 1781 | 23700 | 4.27 | 8.9 | 0.00 | 5 | -0.92 | +0.02 |
| 1810 | 21040 | 4.11 | 7.7 | 0.35 | 24 | -0.49 | +0.45 |
| 1820 | 19370 | 4.05 | 7.0 | 0.45 | 14 | -0.95 | -0.01 |
| 1840 | 21300 | 4.26 | 7.4 | 0.00 | 11 | -0.93 | +0.01 |
| 1848 | 18900 | 4.23 | 6.1 | 0.11 | 25 | -0.92 | +0.02 |
| 1861 | 25300 | 4.11 | 10.7 | 0.30 | 10 | -1.12 | -0.18 |
| 1880 | 25400 | 4.21 | 10.3 | 0.08 | 72 | -0.83 | +0.11 |
| 1886 | 23300 | 4.11 | 9.2 | 0.32 | 13 | -1.16 | -0.22 |
| 1923 | 21400 | 3.78 | 9.5 | 0.79 | 17 | -0.88 | +0.06 |
| 1933 | 24090 | 4.14 | 9.7 | 0.26 | 66 | -0.77 | +0.17 |
| 1950 | 23100 | 4.13 | 9.0 | 0.28 | 34 | -0.95 | -0.01 |
| 2058 | 21720 | 4.24 | 7.7 | 0.06 | 21 | -0.96 | -0.02 |
| 2205 | 19390 | 3.74 | 8.1 | 0.83 | 9 | -0.90 | +0.04 |
| 2344 | 19420 | 3.76 | 8.0 | 0.81 | 61 | -0.85 | +0.09 |
| 2373 | 20400 | 3.92 | 7.9 | 0.65 | 57 | -1.08 | -0.14 |
| 2517 | 16600 | 3.31 | 8.1 | 1.02 | 70 | -0.54 | +0.40 |
| 2618 | 22900 | 3.39 | 14.4 | 1.01 | 47 | -0.60 | +0.34 |
| 2633 | 18000 | 3.38 | 8.8 | 1.02 | 18 | -0.75 | +0.19 |
| 2688 | 19200 | 3.86 | 7.3 | 0.73 | 17 | -1.02 | -0.08 |
| 2756 | 16590 | 4.06 | 5.4 | 0.46 | 26 | -1.09 | -0.15 |
| 2824 | 20210 | 3.86 | 8.4 | 0.72 | 12 | -0.97 | -0.03 |
| 2928 | 23650 | 3.83 | 11.2 | 0.73 | 26 | -0.78 | +0.16 |
| 3023 | 22400 | 3.86 | 9.7 | 0.71 | 42 | -0.89 | +0.05 |
| 7426 | 16540 | 3.60 | 6.8 | 0.93 | 29 | -0.94 | 0.00 |
| 7862 | 17750 | 4.19 | 5.7 | 0.17 | 34 | -1.09 | -0.15 |
| 7929 | 16700 | 3.64 | 6.7 | 0.91 | 40 | -0.94 | 0.00 |
| 7996 | 15980 | 3.60 | 6.5 | 0.92 | 35 | -0.97 | -0.03 |

TABLE 2. (Cuntinue)

| 1 | 2 | 3 | 4 | 5 | 6 | 7 | 8 |
|---|---|---|---|---|---|---|---|
| 8243 | 24300 | 3.32 | 18.0 | 1.01 | 67 | -0.17 | +0.77 |
| 8385 | 15360 | 3.76 | 5.6 | 0.82 | 19 | -0.95 | -0.01 |
| 8439 | 17400 | 3.31 | 8.8 | 1.02 | 19 | -1.06 | -0.12 |
| 8549 | 20400 | 3.97 | 7.8 | 0.60 | 8 | -0.96 | -0.02 |
| 8768 | 18090 | 3.94 | 6.5 | 0.65 | 8 | -1.02 | -0.08 |
| 9005 | 22340 | 3.79 | 10.5 | 0.76 | 15 | -0.75 | +0.19 |

The next to last column of Table 2 gives the value of $\log(N/O) = \log\varepsilon(N) - \log\varepsilon(O)$. It should be noted that for 22 of the stars in Table 2 the values of $T_{eff}$, $\log g$, and $\log(N/O)$ are taken from Ref. 13, as opposed to the other stars for which data from Ref. 12 were used. Those values are more refined than the data of Ref. 12 for the same stars. The abundances of N and O for these stars differ somewhat between Refs. 13 and 12, but it is important that the ratio N/O is the same within the limits of error. The last column of Table 2 lists the value of [N/O] = $\log(N/O) + 0.94$, the difference between the value of $\log(N/O)$ found for each star and the initial value of $\log(N/O)$ at the beginning of the main sequence. According to Table 1, the latter is given by $\log(N/O) = -0.94$ for the stars in the same sample. It is interesting that the same initial value of $\log(N/O) = -0.94$ is assumed for the model calculations of Ref. 7 which are used in the following.

Table 3 lists the parameters of six B-stars with effective temperatures $T_{eff} > 26000$ K [12]. As noted above, for these hot stars the abundances of N and O may be reduced because of overionization, but it is to be expected that the ratio N/O will not be influenced significantly by this effect. Nevertheless, these six stars will be examined separately in the following analysis of N/O (see Figs. 1 and 2). It turns out that they are fully consistent with the behavior of the stars in Table 2.

TABLE 3. Parameters of Six B-stars with Temperatures $T_{eff} > 26000$ K

| HR | $T_{eff}$ K | $\log g$ | $M/M_\odot$ | $t/t_{MS}$ | $V\sin i$ km/s | $\log(N/O)$ | [N/O] |
|---|---|---|---|---|---|---|---|
| 1756 | 27900 | 4.22 | 12.4 | 0.04 | 33 | -1.01 | -0.07 |
| 1855 | 30700 | 4.42 | 15.0 | 0.00 | 20 | -0.84 | +0.10 |
| 1887 | 27500 | 4.13 | 12.5 | 0.24 | 30 | -0.97 | -0.03 |
| 2479 | 28500 | 3.65 | 18.8 | 0.80 | 90 | -0.31 | +0.63 |
| 2739 | 29900 | 4.10 | 15.1 | 0.27 | 47 | -0.59 | +0.35 |
| 8797 | 27200 | 3.98 | 13.1 | 0.51 | 47 | -0.58 | +0.36 |

It should be noted that the accuracy of the values of [N/O] given in Table 2 averages ±0.14 dex, although for individual stars the error reaches ±0.20 dex. The error in determining [N/O] for the six stars in Table 3 is somewhat higher: for four of the stars it averages ±0.16 dex and for two of the stars (HR 2479 and HR 2739) it is ±0.24 and ±0.25 dex. The comparatively large errors in determining [N/O] for HR 2479 and HR 2739 are explained by the lower accuracy for the nitrogen abundances of these stars. The errors in [N/O] for HR 2479 and HR 2739 are indicated in Figs. 1 and 2.

The evolution of the observed values of [N/O] over the MS stage should, in principle, depend on three parameters: the mass, age, and rotation velocity of a star. Thus, general information on these parameters for the stars in Tables 2 and 3 is of interest. The masses M of the stars in Table 2 range from 5.4 to 18.0 $M_\odot$, while Table 3 contains only relatively massive stars with M ranging from 12.4 to 18.8 $M_\odot$. The relative age $t/t_{MS}$ of these stars ranges from zero (the start of the MS) to $t/t_{MS}$ = 1.01-1.02. Recall that a value of $t/t_{MS}$ = 1 corresponds to a star at the time of the end of the MS stage, i.e., the stars with $t/t_{MS}$ = 1.01-1.02 have just finished this stage. The observed rotation velocities for all the stars $V\sin i$ < 100 km/s. (Note that the lists [12,13] of stars used in this paper did not include stars with higher $V\sin i$.) It should be emphasized that for some stars the actual rotation velocity V at the equator can be considerably higher than the projection $V\sin i$ of V on the line of sight.

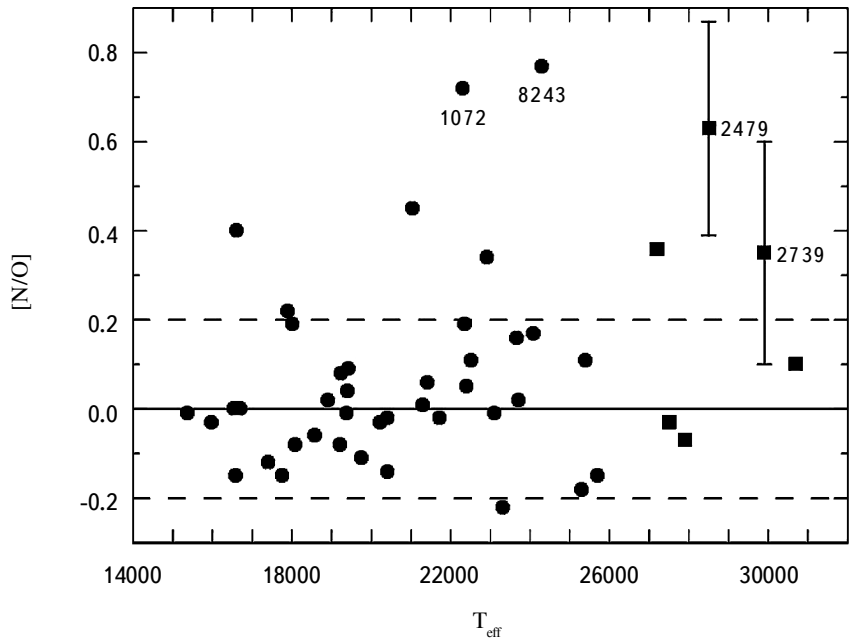

Fig. 1. *[N/O]* as a function of effective temperature $T_{eff}$. The solid circles correspond to stars with $T_{eff}$ < 26000 K from Table 2 and the solid squares, to stars with $T_{eff}$ > 26000 K from Table 3. The errors in determining *[N/O]* are indicated for the stars HR 2479 and HR 2739 (in those two cases the errors are especially large).

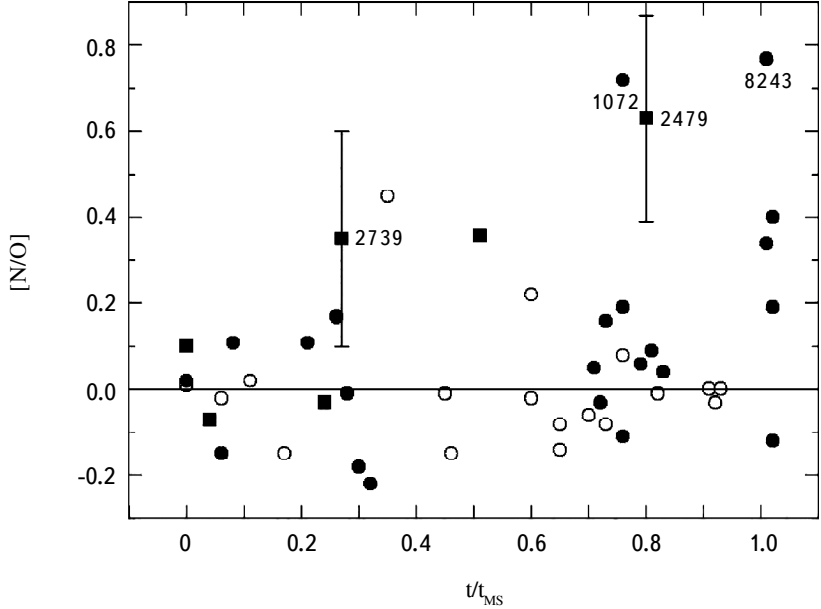

Fig. 2. *[N/O]* as a function of relative age $t/t_{MS}$. The solid circles correspond to the stars of Table 2 with masses $M \geq 8M_\odot$ and the open circles, to stars with $M < 8M_\odot$. The solid squares correspond to the stars from Table 3. The data for the stars HR 2739, 1072, 2479, and 8243 are indicated in the figure. The errors in determining *[N/O]* are indicated for HR 2739 and HR 2479.

## 4. Dependence of the ratio N/O on effective temperature

These data on [N/O] can be analyzed initially by plotting *[N/O]* as a function of effective temperature $T_{eff}$. This dependence is plotted in Fig. 1 for $T_{eff}$ ranging from 15360-30700 K. Here the zero line corresponds to the initial ratio N/O at the start of the MS. In terms of the theory, stars with an initial rotation velocity $V_0 = 0$ km/s, for which the abundances of N and O (and, thereby, the ratio N/O) do not vary from the start to the end of the MS stage, should lie on this line.

The obvious conclusion that follows from Fig. 1 is that most of the MS B-stars studied here over the entire wide range of $T_{eff}$ have *[N/O]* lying close to the zero line; the spread of ±0.2 dex about that line is comparable to the error in the determination of *[N/O]*. (A band of width ±0.2 dex about this line is shown in Fig. 1.) Only 8 of the 46 stars, i.e., 17%, had values of *[N/O]* significantly greater than 0.2 (ranging from 0.34 to 0.77). As for the other 38 stars (83 %), it appears that they, with low values of $[N/O] \approx 0$, also had low initial rotation velocities $V_0$. In the following, it will be shown that current studies confirm this point.

The other conclusion that follows from Fig. 1 is that [N/O] does not have any particular trend with $T_{eff}$. As expected, the stars with $T_{eff}$ > 26000 K in Table 3 (squares) have no systematic differences in *[N/O]* from the stars with $T_{eff}$ < 26000 K in Table 2 (solid circles). This last conclusion is confirmed by comparing *[N/O]* with other parameters (see below).

**5. The dependence of the ratio N/O on age**

Figure 2 shows the variation in *[N/O]* with the relative age of the stars $t/t_{MS}$. In order to distinguish the stars in terms of mass *M*, different data points are used to indicate the stars with $M \geq 8 M_\odot$ and $M < 8 M_\odot$ from Table 2. Special points (squares) are used for the six more massive stars with $M = 12.4 - 8.8 M_\odot$ from Table 3. Here the zero line, as in Fig. 1, corresponds to the initial ratio N/O at the start of the MS.

Most of the B-stars in Fig. 2 lie near the zero line within a spread of ±0.2 dex, regardless of their relative age $t/t_{MS}$. Recall that this behavior was noted in the discussion of Fig. 1. The reason that $[N/O] \approx 0$ for these stars is related to their rather low initial rotation velocity $V_0$ (see below).

Figure 2 shows that the ten stars with ages $0 \leq t/t_{MS} < 0.25$, i.e., which are found at the very start of evolution in the MS, all have $[N/O] \approx 0$. Higher values of *[N/O]* are observed only in stars which have evolved with ages $t/t_{MS} \geq 0.3$. An especially large spread in *[N/O]* is found in Fig. 2 for five stars which have just finished the MS stage (for them, $t/t_{MS}$ = 1.01-1.02); here *[N/O]* ranges from -0.12 for HR 8439 to +0.77 for HR 8243.

It is interesting to note that the star HR 8243, which has the largest *[N/O]* = 0.77, also has almost the largest mass $M = 18 M_\odot$ of the stars studied here. Two of the other stars in Fig. 2 which also had high values of *[N/O]* = 0.6-0.7 and are also close to completion of the MS stage, HR 1072 and HR 2479, also have high masses $M$ = 9.8 and 18.8 $M_\odot$. It will be shown below that the higher values of *[N/O]* = 0.4-0.8 are consistent with models of stars with initial rotation velocities $V_0$ = 200-300 km/s.

In concluding our discussion of Fig. 2, it should be noted that, as expected, the six stars with $T_{eff}$ > 26000 K from Table 3 are fully consistent with the general pattern for the stars with $T_{eff}$ < 26000 K from Table 2.

**6. The dependence of the ratio N/O on rotation velocity**

According to the theory, one other parameter, the rotation velocity, should play an important role in the evolution of *[N/O]* in the early B-stars. Figure 3 is a plot of the found values of *[N/O]* as a function of the observed rotation velocity *V*sin*i*, the projection of the rotation velocity *V* at the equator on the line of sight. For all the stars in Fig. 3, *V*sin*i* < 100 km/s, but, as noted above, the actual rotation velocity *V* at the equator for some of the stars may be considerably higher than *V*sin*i*.

In order to account for the dependence of *[N/O]* on the other two parameters (mass *M* and age $t/t_{MS}$), Figure 3 is shown in two frames. In the top frame (Fig. 3a) the stars with masses $M \geq 8 M_\odot$ and $M < 8 M_\odot$ are indicted by different plot points. In the bottom frame (Fig. 3b) stars with ages $t/t_{MS} \leq 0.3$ (the start of the MS) and $t/t_{MS} > 0.5$ (second half of life on the MS) are indicated by different points. Some stars with $t/t_{MS}$ between 0.3 and 0.5 are not shown in Fig. 3b.

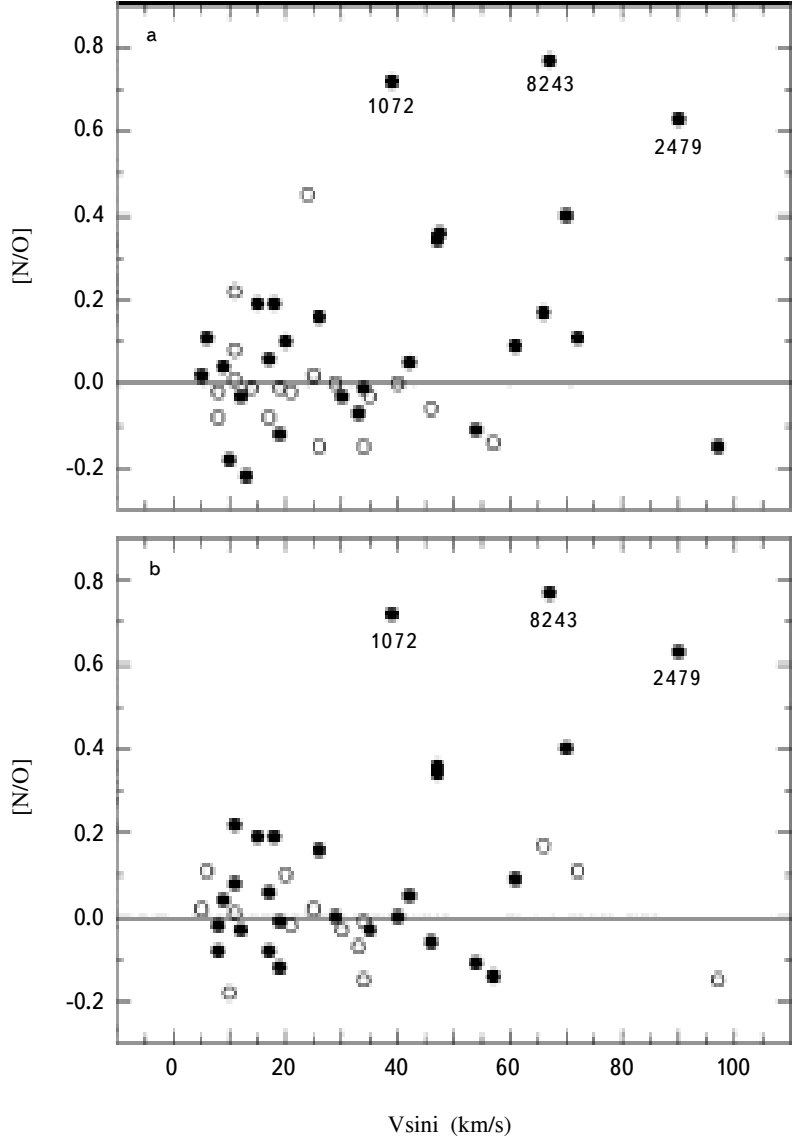

Fig. 3. *[N/O]* as a function of the observed rotation velocity *V*sin*i*. In the top frame (Fig. 3a) the stars are separated by mass: objects with masses $M \geq 8 M_\odot$ are plotted as solid circles and those with $M < 8 M_\odot$, as hollow circles. In the lower frame (Fig. 3b), the stars are separated by relative age: the solid circles correspond to objects with ages $t/t_{MS} > 0.5$ and the hollow circles, to objects with $t/t_{MS} \leq 0.3$.

In Figs. 1 and 2 the six stars with $T_{eff} > 26000\,K$ from Table 3 did not manifest any systematic differences in their values of *[N/O]* from the stars with $T_{eff} < 26000\,K$ from Table 2. It turns out that in the plots of *[N/O]* as a function of *Vsini* in Fig. 3, these stars fit the general pattern well, so they are not indicated by special plot points in Fig 3 (so as to avoid complicating the figure). (The same comment applies to the last figure, Fig. 4.)

As already noted, rotation model calculations [7] have been done for masses $M$ ranging from 15 to 1.7 $M_\odot$ and initial rotation velocities $V_0$ ranging from zero to 0.95 times the critical velocity. These results [7] imply that for a rotation velocity $V_0 = 40$ km/s, even in a massive star with $M = 15 M_\odot$, *[N/O]* does not change by the end of the MS. A velocity $V_0 > 100$ km/s is needed for a detectable rise in *[N/O]*. For example, Ref. 7 shows that for $V_0 = 130$ km/s, *[N/O]* increases to 0.33 dex for $M = 15 M_\odot$ and only to 0.20 dex for $M = 12 M_\odot$.

This raises the following question: what are the actual rotation velocities observed in early MS B-stars? It turns out that the observed values of *Vsini* for many stars are, in fact, small. In particular, according to Ref. 17, the maximum in the rotation velocity distribution for MS dwarfs in classes B0-B2 lie within the range *Vsini* = 0-20 km/s, although the entire range of values of *Vsini* extends to *Vsini*~400 km/s. A similar result was obtained earlier [18]: the maximum in the distribution of *Vsini* for early MS B-stars lies in the range of 0-50 km/s. In a study of the rotation velocity in MS B-stars, Abt, et al. [19], found that a fairly large fraction of these stars have an initial rotation velocity at the equator of $V_0$~50 km/s.

Given these remarks, we can now return to Fig. 3 and discuss some of its features. First, most of the stars (32 of 47, i.e., 68%) have low rotation velocities *Vsini* = 5-40 km/s. This is because the actual fraction of early B-stars with $V \leq 40$ km/s is, in fact, large (see above). Second, the overwhelming majority of these stars have *[N/O]* near the zero line in Figs. 3a and 3b. This means that the values of *Vsini* for them are fairly close to the actual rotation velocities *V*, while *[N/O]* for these stars should not change along the MS [7], regardless of their mass and age. For the pair of stars (HR1072 and HR1810) with $V\sin i < 40$ km/s which have a high value of *[N/O]* in Figs. 3a and 3b, we can assume that the actual velocity *V* is much higher than $V\sin i$.

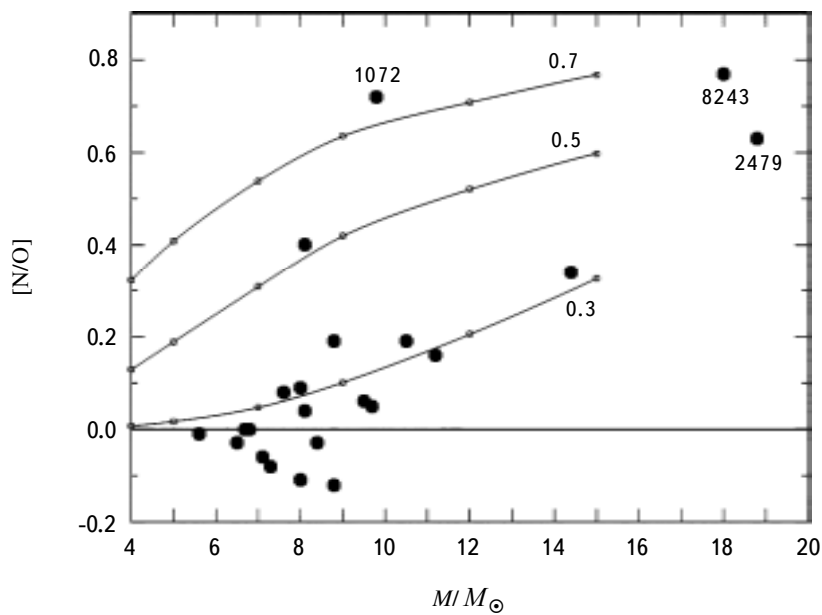

Fig. 4. *[N/O]* as a function of mass $M$ for B-stars at the end of the MS stage ($t/t_{MS}$ = 0.70-1.02). The smooth curves are theoretical curves based on model calculations [7] for three values of the initial angular velocity of rotation $\Omega/\Omega_{crit}$ = 0.3, 0.5 and 0.7.

Furthermore, the stars in Figs. 3a and 3b with higher rotation velocities $V \sin i \geq 40$ km/s have a large observed scatter in *[N/O]*. On one hand, all the elevated values of *[N/O]* > 0.3 in Fig. 3a belong to stars with relatively high masses $M \geq 8 M_\odot$. On the other hand, all of these stars with *[N/O]* > 0.3 (except one) have the same relative age $t/t_{MS} > 0.5$ (Fig. 3b); that is, they are approaching the end of the MS stage. In terms of the theory, these results are fully to be expected.

## 7. The dependence of the ratio N/O on mass

In order to examine the dependence of *[N/O]* on mass M and then to make a quantitative comparison of the results with theoretical predictions, of the B-stars studied here we shall select those with $t/t_{MS} \geq 0.7$, i.e., those which are close to completing the MS stage. *[N/O]* is plotted as a function of *M* for these stars (22 with masses M from 5.6 to 18.8 $M_\odot$) in Fig. 4. Only one star from Table 3 (HR 2479) appears in this figure; all the others are taken from Table 2. Here the smooth curves show the results of calculations [7] for masses $M = 4-15 M_\odot$ and three initial rotation velocities $\Omega/\Omega_{crit} = 0.3$, 0.5, and 0.7, where is the angular rotation velocity and $\Omega_{crit}$ is the critical rotation velocity (the calculations [7] were done for all values of $_{crit}$ from 0 to 0.95).

$V_0$, the initial linear rotation velocity at the equator, depends on the mass *M* for a constant value of $\Omega/\Omega_{crit}$. For $\Omega/\Omega_{crit} = 0.3$, $V_0$ ranges from 116 km/s for $M = 5 M_\odot$ to 132 km/s for $M = 15 M_\odot$ [7]. For $\Omega/\Omega_{crit} = 0.5$, the corresponding change in $V_0$ is from 192 to 241 km/s, and for $\Omega/\Omega_{crit} = 0.7$, from 277 to 333 km/s.

Figure 4 shows that for most of the stars shown here, which have comparatively low $[N/O] \leq 0.3$, the most appropriate models from a theoretical standpoint are with $\Omega/\Omega_{crit} = 0$ and $\Omega/\Omega_{crit} = 0.3$, which correspond to initial rotation velocities $V_0$ from 0 to 130 km/s. For the four stars with the highest values of *[N/O]* = 0.40-0.77, the models with $\Omega/\Omega_{crit} = 0.5$ ($V_0 \approx 220$ km/s) and $\Omega/\Omega_{crit} = 0.7$ ($V_0 \approx 300$ km/s) fit well. We note that the highest *[N/O]* = 0.8-0.9 are predicted in the calculations [7] for a model of $M = 15 M_\odot$ with rotation velocities $\Omega/\Omega_{crit} = 0.80 - 0.95$ ($V_0 \approx 400 - 500$ km/s), but no such values of [N/O] appear in Figs. 1-4.

Thus, an analysis of the values of *[N/O]* for early MS B-stars with M ranging from 5 to 19 $M_\odot$ yields the following conclusions. Since most of these stars are observed to have rotation velocities $V_0 <$ 100 km/s in the beginning of the MS, in full agreement with the theoretical models the observed $[N/O] \leq 0.2$ at the end of the MS; that is, the ratio N/O does not increase significantly during the MS stage. Only for initial velocities $V_0$ = 200-300 km/s will high values of [N/O] = 0.4-0.8 occur at the end of the MS.

In connection with these conclusions there is some interest in the next stage of post-MS evolution into which the early B-stars with masses $M = 4 - 20 M_\odot$ enter. This is the stage of A-, F-, and G-supergiants and bright giants. It is known that when a certain effective temperature ($T_{eff} < 5900$ K) is reached in these stars, deep convective mixing sets in; as a result, the anomalies in the atmospheric abundances of C, N, and O that show up during the MS stage can become considerably stronger. For those AFG-supergiants and giants that have still not reached the deep convective mixing phase, the C, N, and O anomalies acquired during the MS are retained.

The non-LTE abundances of C and N, as well as the ratio C/N in the atmospheres of 36 galactic AFG-supergiants and bright giants, have been analyzed [5]. A comparison with rotating star models showed that these stars have initial rotation velocities $V_0$ from 0 to 300 km/s. If these stars are treated as post-MS objects (i.e., prior to the deep convective mixing phase), then the largest C and N anomalies correspond to a velocity $V_0$ = 200-300 km/s. Thus, the conclusions of this paper are fully consistent with those of Ref. 5 for colder stars with the same masses in a later stage of evolution.

## 8. Conclusion

The major results of this paper are the following:

1. It has been shown that for early MS B-stars, of the three ratios N/C, C/O, and N/O regarded as indicators of stellar evolution, only N/O can be considered to be reliable, since it is insensitive to possible overionization of NII and OII ions. On the other hand, for stars with effective temperatures $T_{eff}$ > 18500 K, N/C and C/O, which involve carbon, may contain systematic errors owing to neglect of overionization of CII ions.

2. The value of [N/O], the ratio N/O normalized to the initial value (on a logarithmic scale), has been determined for 46 early MS B-stars with masses M ranging from 5.4 to 18.8 $M_\odot$. The dependence of *[N/O]* on such parameters as the effective temperature $T_{eff}$, relative age $t/t_{MS}$, observed rotation velocity $V\sin i$, and stellar mass *M* has been analyzed.

3. For most of the stars studied here, $[N/O] \approx 0$ with a spread of ±0.2 dex; this result does not depend at all on the parameters $T_{eff}$, $t/t_{MS}$, $V\sin i$, and *M*. In other words, these stars manifested no significant changes in the ratio N/O during the MS stage. Only 8 stars out of the 46 (17%) had values of *[N/O]* much higher than the possible error (from 0.34 to 0.77).

4. A comparison has been made with rotating star model calculations [7]. It was found that there are two reasons for the large number of stars with $[N/O] \approx 0$: first, according to the calculations [7] for an initial rotation velocity $V_0$ < 100 km/s the ratio N/O changes little by the end of the MS stage (*[N/O]* < 0.2) and second, a majority of the early MS B-stars do, in fact, have low initial rotation velocities $V_0$.

5. The early MS B-stars with elevated values of *[N/O]* = 0.4-0.8 can be explained by models with initial rotation velocities $V_0$ = 200-300 km/s. Estimates of this kind with $V_0$ ranging from 0 to 200-300 km/s have been obtained [5] in an analysis of the abundances of C and N and the ratio C/N for stars with the same masses in the next stage of evolution, with A-, F-, and G-supergiants and bright giants.